# Central universal force field to explain solar orbital radial acceleration and other universal phenomena

Kamal Barghout

Ag & Bio Eng. Department, University of Saskatchewan, 57 Campus Drive, Saskatoon, SK S7N 5A9 CANADA

**Abstract:** I investigate a repulsive central universal force field on the behavior of celestial objects. I show its negative tidal effect on the solar orbits as experienced by Pioneer spacecrafts. I explain several cosmological effects in light of this force.

## 1. Introduction

Attempts to employ Newtonian mechanics to explain various cosmological anomalies such as Pioneer anomaly have failed so far. A vector field originates from the center of the universe (assuming a spherical universe), possibly local universe, could solve many cosmological problems. This vector field could be the work of a constant repulsive force that could be the one responsible for the initiation of the Big Bang event and situated at the center of the universe or it could be the result of an existing mass-repulsive entity distributed throughout the cosmos (suggested candidate is anti-mass). The result of this field will be acceleration in the radial direction which presents a strong candidate to "Dark Energy". Previous modeling of repulsive-mass-antimass universe [1-3] failed to explain much of cosmic behavior. In this paper I present a model of the universe with equal amounts of mass and antimass repelling each other, reserving a universal radial gravitational field. Our Milky Way galaxy and the galactic components including the solar system will feel the radial acceleration. A negative-tidal force resulting from a component of the repulsive central force acting on the solar system will result in a sunward acceleration felt by the sun orbiters. I will explain several of the cosmological anomalies mainly by considering the action of the central force. I will demonstrate that the success of Newtonian mechanics in situations like our solar system can be extended beyond the local system to explain the rotation speed of stars in galaxies (galactic rotation curves) using the proposed universal force based on the assumption that Newtonian gravity is a good approximation without the need for a large amount of dark matter to be present.

## 2. Effect of a repulsive universal force on galactic and solar system dynamics

Consider a situation where the orbit of the planet is in the radial direction of the universe so the planet would be between the center of the universe and the sun at one point and in the opposite side after half orbital period. The proposed universal repulsive force will have a negative tidal effect on the solar system and can cause inward bulge (toward the sun) of the planet's orbit as acceleration of the sun in the upward direction is larger than that of the planet when the planet is situated above the sun relative to the center of the universe while the acceleration of the planet is larger than that of the sun when it is situated below the sun (or component of the repulsive force toward the sun when the orbital plane is not at $90^0$ with the radial direction). Fig. 1 shows the force acting on a celestial system with a plane perpendicular to the universal radial direction.

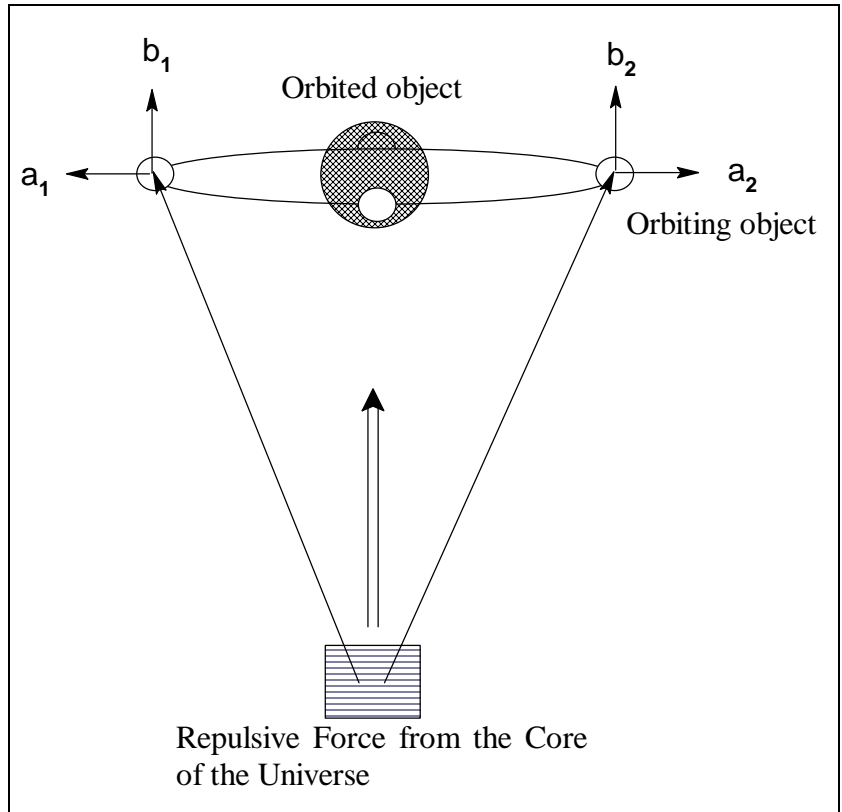

Fig.1. Repulsive force acting on a celestial system with a plane perpendicular to the universal radial direction (not to scale).

The repulsive force also affects the dynamics of galaxies. The unequal forces of $b_1$ and $b_2$ in an orbital plane that is not perpendicular to the universal radial direction as might be the case of the Milky Way galaxy will impart unequal accelerations and an eventual perpendicular orbital orientation (tangential to a spherical shell concentric with the center of the universe). This produces an oscillation of the galactic plane about the tangential plane because the produced net torque will reverse in each side when the galactic plane overshoots the tangential plane and eventually alignment of the galactic plane with the tangential plane will occur at some point. This will be pronounced with each individual star as its orbital plane oscillates about the tangential plane.

While the perpendicular components of the repulsive forces (b-forces in fig. 1) initiate the oscillation, the in-plane forces (a-forces in fig. 1) act as the restoring forces. Celestial systems will be oscillators oriented at random relative to each other due to the universal repulsive force acting on them.

The sun-orbiting objects, as in the case of the Pioneer spacecraft, feel negative tidal forces in all directions due to their orbital plane tilting relative to the tangential plane with varying magnitudes toward the orbited object as opposed to that of the normal positive tidal effect as in the case of the moon and the earth.

## 3. Pioneer anomaly

Pioneer anomaly is a constant acceleration directed towards the Sun of $a = (8.74 \pm 1.33) \times 10^{-10} \, m \, s^{-2}$ registered by Pioneer spacecrafts [4, 5]. Attempts to explain this anomaly have failed so far.

Following the previous argument in section 2 we can explain the Pioneer Effect, which could be due to an external origin such as the action of the universal repulsive force acting on the Pioneer spacecrafts to produce negative tidal acceleration, which resulted in a constant sunward acceleration.

## 4. Modeling the sun's dynamics as simple harmonic motion

The sun oscillation about the galactic plane can be treated as a simple harmonic motion modeled by a simple pendulum that extends from the center of the galaxy to the sun with the sun acting as the bob and a component of the repulsive central universal force as the restoring force. Using the period of the simple pendulum as the oscillation period of the sun to compute the gravitational field strength at the sun location when the galactic plane is perpendicular to the universal radial direction, $P = 2\pi \sqrt{\dfrac{l}{g}}$, with the distance of the sun from the galaxy center is about 27700 *ly* and the oscillation period is about 66 M years about the galactic plane, g (The gravitational field strength in the immediate neighborhood of the sun, which originates from the repulsive force) is about $0.24 \times 10^{-10} \, m \, s^{-2}$. This should give a restoring force of the sun-galaxy pendulum of $0.24 \times 10^{-10} \, M_{sun}$.

## 5. Maximum height of stars from the galactic plane

Models of gravitational force law perpendicular to the galactic plane allowed the determination of the galactic velocity-distance relation with close fit to the velocity-distance observations. The velocity-distance obtained from the simple harmonic modeling of the sun conforms to observation.

Fig. 2 illustrates the mechanism of the sun vertical oscillation about the galactic plane as modeled by a simple harmonic motion.

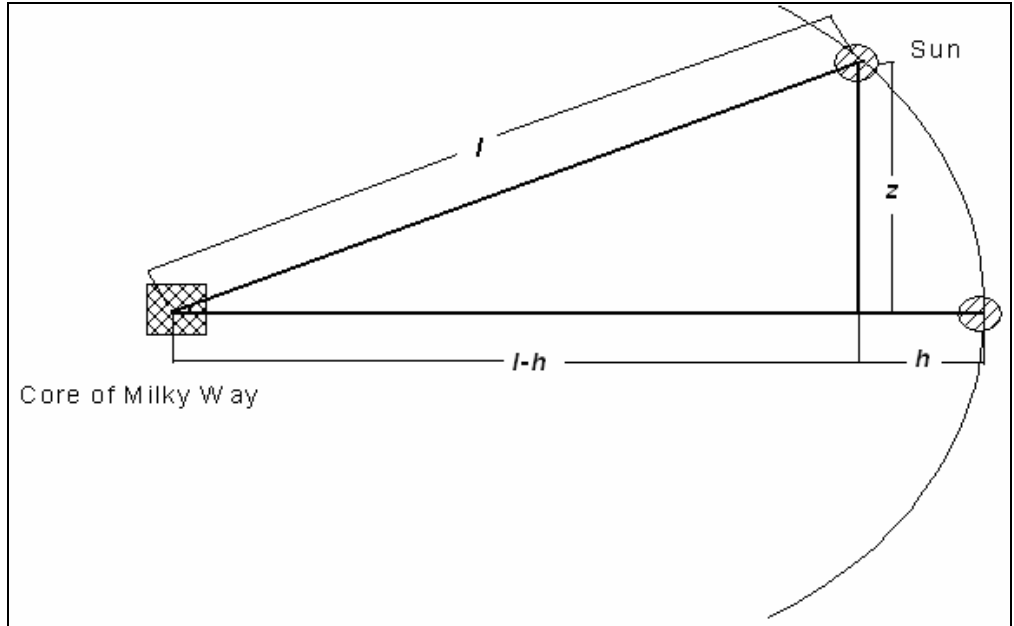

Fig. 2. Vertical oscillation of the sun modeled as simple harmonic motion.

Table 1 shows the sun galactic velocity-distance computed from the simple harmonic modeling compared with that of [6] computed from distribution function modeling technique.

Table 1. Sun galactic velocity-distance.

| Velocity m/s at z=0 | $Z_{maximum}$ (Pc), this paper (simple harmonic model) | $Z_{maximum}$ (Pc), Kuikken & Gilmore paper |
|---|---|---|
| 0 | 0 | 0 |
| 8400 | 149 | 100 |
| 15700 | 279 | 200 |
| 21900 | 390 | 300 |
| 27300 | 486 | 400 |
| 32100 | 571 | 500 |
| 36500 | 650 | 600 |
| 40600 | 722 | 700 |
| 44500 | 792 | 800 |
| 48300 | 859 | 900 |
| 51800 | 921 | 1000 |
| 58600 | 1042 | 1200 |
| 68300 | 1214 | 1500 |
| 83200 | 1477 | 2000 |

In table 1 the following governing equations were used to compute the velocity-distance of the sun following the illustration in fig. 2.

$$v^2 = 2gh \tag{1}$$

$$l^2 = z^2 + (l-h)^2 \tag{2}$$

Where g is the gravitational field strength at the sun vicinity and taken as $a = 8.74 \times 10^{-10} \, m\,s^{-2}$ and *l* is the distance of the sun-galaxy center taken as 27700 *ly*.

## 6. Galactic Warp

Galactic warp could be explained as well in terms of the central repulsive force of the universe. The warp of galaxies is normally explained in the literature in terms of contained dark matter. In this paper dark matter is explained in terms of the central repulsive force so is the galactic warp. Fig. 3 shows the repulsive force acting on a proposed galaxy. If the galactic plane is not tangential to a spherical shell with its center as that of the universe, the central repulsive force will act with different magnitudes of perpendicular force components along the galaxy. The longest arm of force will provide the smallest component while the shortest will provide the largest component. Since the whole galaxy is accelerated in the radial direction due to the central repulsive force but one end of the galaxy is accelerated faster than the opposite one, with perpendicular force distribution as shown in Fig. 3, this will produce a warp of the galaxy.

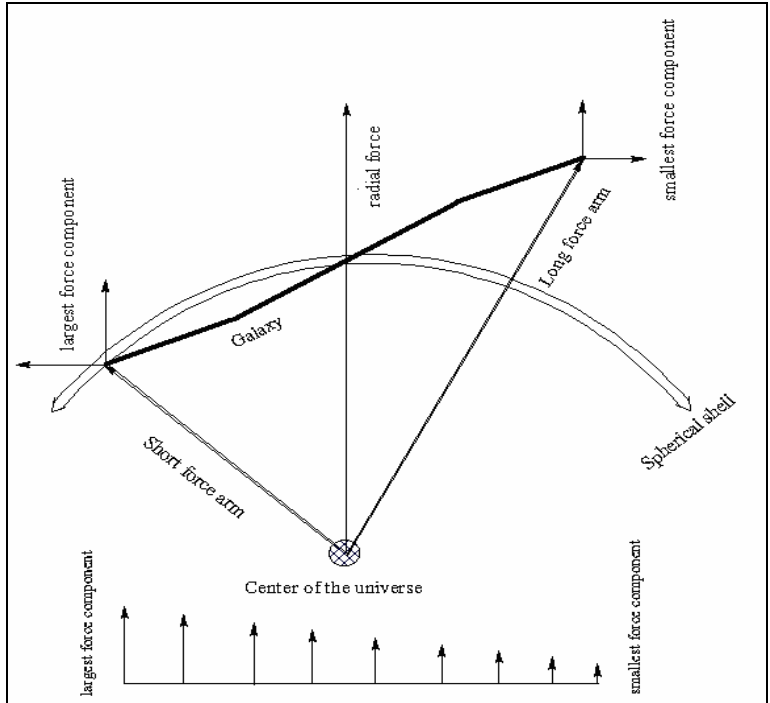

Fig. 3. Galactic warp (not to scale).

## 7. Effect of repulsive force on Milky Way Galaxy at the sun vicinity (dark matter halo)

Rotation curve of our galaxy shows that the sun is moving about 60 km/s faster than what is contributed by visible matter as the actual sun's velocity measures 220 km/s while that contributed by visible matter is about 160 km/s. The centripetal acceleration of a negative tidal force component on the sun possibly contributed by the proposed central negative force can account for the difference. Centripetal acceleration of $0.24 \times 10^{-10}\, m\, s^{-2}$ (computed from the pendulum model) will account for the difference; where $V = \sqrt{ar}$, $r$ is $27700\ ly$, V is close to 60 km s$^{-1}$. This is usually interpreted as a dark matter halo.

## 8. Dark matter distribution

It is becoming known to astronomers that the distribution of galaxies follows the distribution of dark matter. It is easy to infer that the distribution of dark matter follows the distribution of baryonic matter by considering an external gravitational field as the cause of baryonic material behavior as discussed above to produce larger orbital velocities of galactic components. This is interpreted by astronomers as the work of an existing dark matter in the cosmos which will only exist where baryonic matter exists.

## 9. Halo density of dwarf galaxies

It is known that smaller galaxies are much more dominated by dark matter. As an example, our Milky Way is a large galaxy with about 50% dark matter. Galaxies with 1/100 of the luminosity of the Milky Way are about 90% dark. The smallest dwarfs known are almost completely made of dark matter. This can be explained by the fact that dark halo is independent of the galaxy size but galactic baryonic mass density is, as dark halo is only dependent on the repulsive gravitational field component. The smaller the galaxy is the less baryonic density it contains. Therefore, for smaller galaxies, dark halo density to baryonic density ratio will be larger.

**10. Modeling a repulsive mass-antimass universe**

The universe can be modeled as one made with equal amounts of mass and antimass and of consecutive matter and antimatter spherical shells with the outermost made from one kind considering that the repelling entity is the antimass and that antimass is attractive to itself. The mechanism of acceleration of the universe arises from the fact that matter (antimatter) of the outermost shell of the universe will feel a net repulsive force due to existence of net opposite mass bounded by the shell. This results in acceleration of the outer shell in the outward direction. The next inner shell (the second shell) faces the same scenario but accelerates at a lesser amount due to the existence of a lesser repulsive force at the proper shell parameters (mass/radius ratio). An observer in a galaxy in the second outermost shell will see galaxies in the outermost shell as well as galaxies in the next inner shell accelerate away from him.

Since the shells of the universe accelerate in the radial direction, galaxies will see accelerations of neighboring same shell galaxies in the shell own curved plane as well as all other possible directions with relative magnitudes. This accelerated movement of galaxies away from apposite kind of galaxies leaves voids [7] among them which accounts for the overall sponge-like structure of the universe. This is the same any where in the universe due to its spherical symmetry. The accelerated movement of galaxies in the context of the universal repulsive force presents a good strong candidate to "Dark Energy".

Analysis of the velocity of a star near the edge of a galaxy that resides in the inner edge of the outer most shell is as follows:

Suppose that the last shell carries antimass and assuming a symmetrical universe, this shell will then enclose a net normal mass with same magnitude. We can now apply Gauss's law to account for the gravitational field that pushes away the galaxy, of galactic plane inclination angle $\theta$ relative to the universal radial direction in the radial direction. The governing equation will be $\int_s g.da = 4\pi G m_{enclosed}$; where $m_{enclosed}$ is the enclosed net normal mass. If we restrict the halo within the volume of the galaxy only, rotational velocity of the star located near the outer edge of the galaxy will be approximated by

$$V = \left(\frac{GM_{galaxy}}{r}\right)^{\frac{1}{2}} + \left(\frac{GM^*}{r}\right)^{\frac{1}{2}} \tag{3}$$

Where $M^*$ is the effective mass due to $m_{enclosed}$, $r$ is the star-galactic center distance and $V$ is a function of $\theta$. Note that the second term in equation (3) accounts for the negative tidal force acting on the star due to the existence of the universal repulsive force as explained earlier.

**11. Rotation curves**

Rotation curves of galaxies (RCs) precisely describe the dynamics of galaxies [8]. RCs are also used to discuss dark matter distribution in halos [9]. It is well known that rotation curves of spiral galaxies are flat. This is usually attributed to the existence of halos surrounding galaxies that contain unseen dark matter, which should in theory constitute the missing mass.

Fig. 4 shows an illustration of the acting force lines on a star with their relative angles.

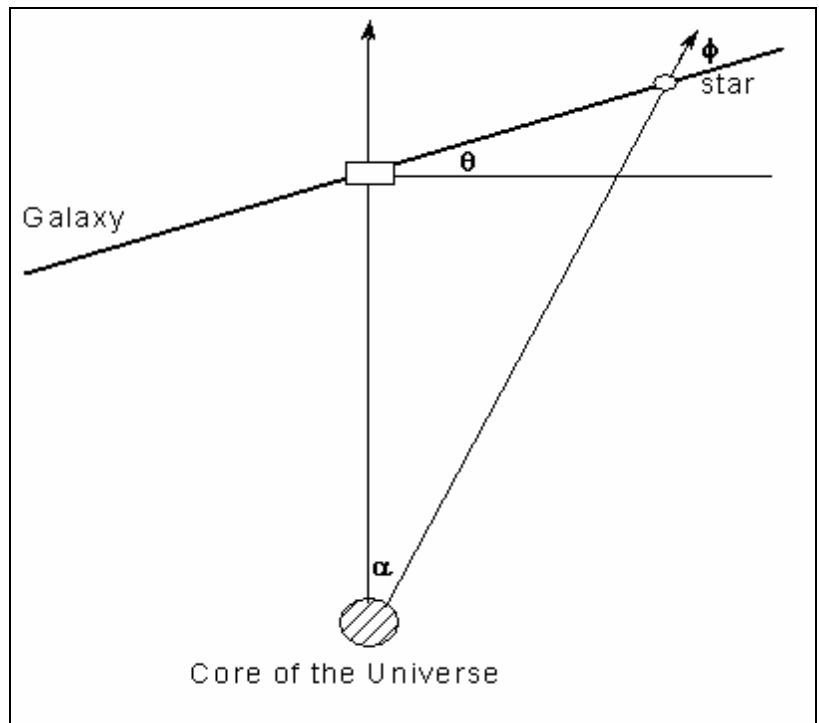

Fig. 4. Universal repulsive force lines on a star with their relative angles (not to scale).

The following discussion shows how this force accounts for the shape of the rotation curve due to dark matter.

If we restrict the halo within the volume of the galaxy only, we can infer from fig. 4 that the repulsive force component of the rotational velocity of the star will be governed by

$$v^2 = d^{-1} \sin^2 \alpha \cos(90 - \theta + \alpha)\left[GM_{enclosed} C^{-2}\right] \quad (4)$$

Where $d$ is the distance from the star to the galactic center; $\phi = 90 - \theta + \alpha$; $v = \sqrt{ad}$; $a = d^{-2} \sin^2 \alpha \cos(90 - \theta + \alpha)\left[GM_{enclosed} C^{-2}\right]$ is the centripetal acceleration; G is the gravitational constant; $M_{enclosed}$ is the net repulsive mass enclosed by the shell; $C$ is the ratio of the shell radius and d; $\theta$ is the inclination angle of the galactic plane with respect to the plane tangential to the shell at the star location and $\alpha$ is the angular separation of the galaxy at the center of the universe as shown in fig. 4. To see how the dark halo's shape looks like we can plot the galactic member's velocity contributed from the repulsive force relative to its corresponding d with $\left[GM_{enclosed} C^{-2}\right]$ as a constant as shown in fig. 5. Here $\theta$ is given an arbitrary value of $0.00048^0$ and $\alpha$ is given arbitrary values in the range of $0^0$ to $0.00048^0$.

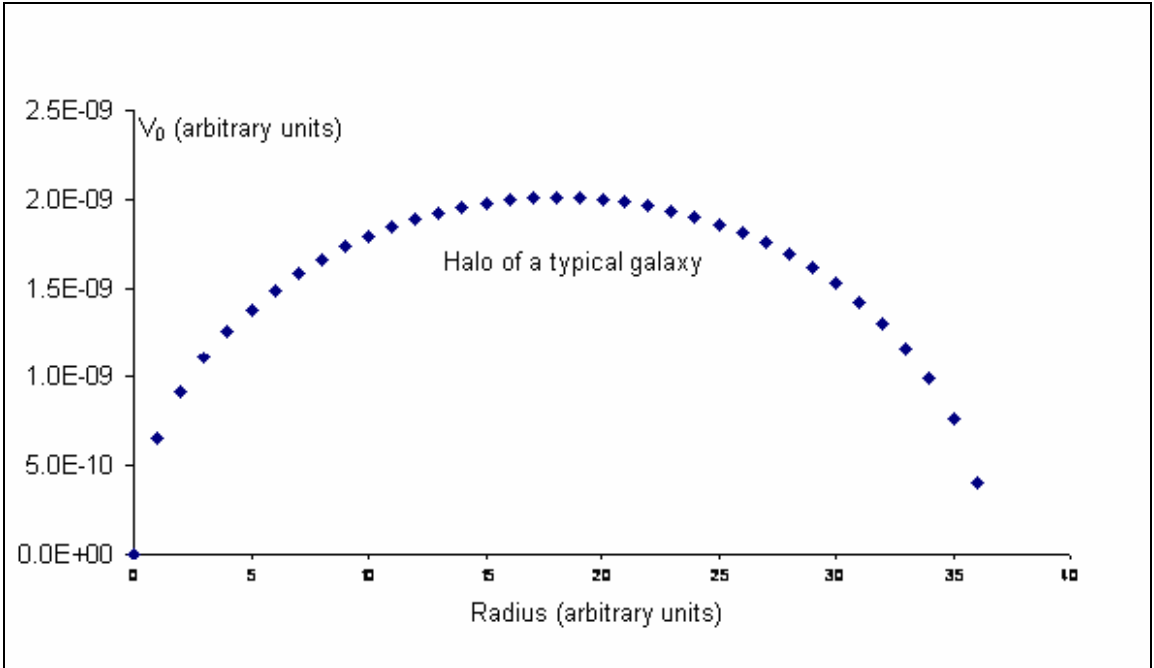

Fig. 5. Typical dark matter halo generated from the planar component of the Universal force: $v^2 = d^{-1} \sin^2 \alpha \cos(90 - \theta + \alpha)\left[GM^* C^{-2}\right]$.

As expected, the increase of the inclination angle will increase the dark halo density as shown in fig. 6 with inclination angle of $10^0$.

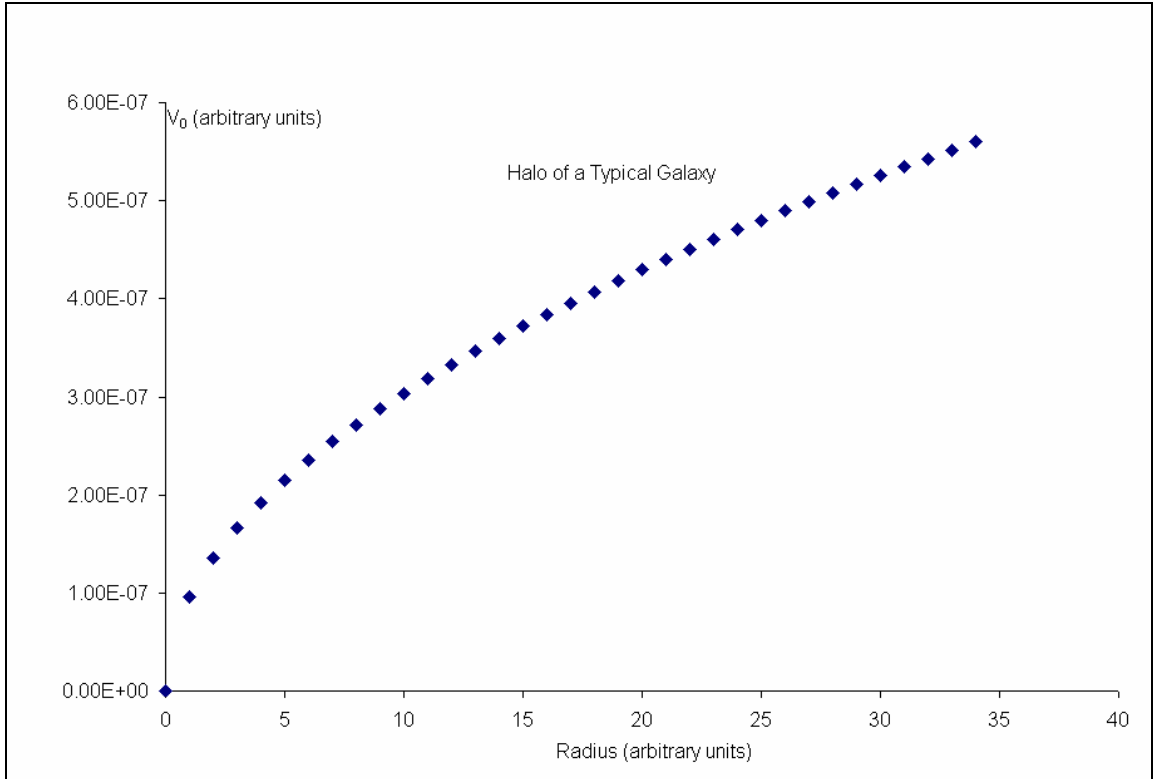

Fig. 6. Shape of dark halo's and its density when the galactic plane is tilted by $10^0$ from a plane perpendicular to the universal radial direction.

The halo's density in general is independent on the baryonic mass density but dependent on the galactic plane orientation (angle $\theta$ in fig. 4).

## 12. Conclusion

I have shown that by accounting for the negative (repulsive) component in Newtonian mechanics we can provide a reasonable explanation to several cosmic anomalies such as dark matter, dark energy, rotation curves, Pioneer anomaly and galactic warp.